\documentclass{article}
\usepackage[utf8]{inputenc}

\usepackage{latexsym}
\usepackage{amssymb,amsmath}
\usepackage{amsthm}
\usepackage{authblk}
\usepackage{graphicx}
\usepackage{sgame}
\usepackage{color}
\usepackage{graphicx}
\usepackage[hidelinks]{hyperref}
\usepackage{empheq}
\usepackage{blkarray}
\usepackage{cancel}
\usepackage{breqn}
\usepackage{comment}
\usepackage{makecell}
\usepackage{tikz} 
\usepackage{setspace}

\usepackage{enumerate}

\usepackage{caption}
\usepackage{subcaption}
\usepackage{booktabs}

\numberwithin{equation}{section}

\newcommand{\bbm}{\begin{bmatrix}}
\newcommand{\bpm}{\begin{pmatrix}}
\newcommand{\ebm}{\end{bmatrix}}
\newcommand{\epm}{\end{pmatrix}}

\usepackage[numbers,sort&compress]{natbib}

\usepackage{graphicx} 

\title{External light schedules can induce nighttime sleep disruptions in a Homeostat-Circadian-Light Model for sleep in young children}
\author[1]{Tianyong Yao}
\affil[1]{Department of Mathematics, University of Michigan, Ann Arbor, MI, USA}
\author[1,2]{Victoria Booth}
\affil[2]{Department of Anesthesiology, University of Michigan, Ann Arbor, MI, USA}
\date{\today}

\newenvironment{authorsummary}
  {\renewcommand{\abstractname}{Author Summary}\begin{abstract}}
  {\end{abstract}}

\begin{document}

\maketitle

\begin{abstract}
Sleep disturbances, particularly nighttime waking, are highly prevalent in young children and can significantly disrupt not only the child’s well-being but also family functioning. Behavioral and environmental strategies, including the regulation of light exposure, are typically recommended treatments for nighttime waking. Using the Homeostatic-Circadian-Light (HCL) mathematical model for sleep timing based on the interaction of the circadian rhythm, the homeostatic sleep drive and external light, we analyze how external light schedules can influence the occurrence of nighttime waking in young children. We fit the model to data for sleep homeostasis and sleep behavior in 2 - 3.5 year olds and identified subsets of parameter ranges that fit the data but indicated a susceptibility to nighttime waking. This suggests that as children develop they may exhibit more or less propensity to awaken during the night. Notably, parameter sets exhibiting earlier sleep timing were more susceptible to nighttime waking. For a model parameter set susceptible to, but not exhibiting, nighttime waking, we analyze how external light schedules affect sleep patterns. We find that low daytime light levels can induce nighttime sleep disruptions and extended bright-light exposure also promotes nighttime waking. Further results suggest that consistent daily routines are essential; irregular schedules, particularly during weekends, markedly worsen the consolidation of nighttime sleep. Specifically, weekend delays in morning lights-on and evening lights-off times result in nighttime sleep disruptions and can influence sleep timing during the week. These results highlight how external light, daily rhythms, and parenting routines interact to shape childrens' sleep health, providing a useful framework for improving sleep management practices.

\end{abstract}

\begin{authorsummary}
Roughly 20-30\% of toddlers and pre-school age children exhibit sleep issues that can include nighttime waking. Recurring sleep disruption can affect cognitive, emotional and behavioral development, thus it is important to understand potential causes and identify treatments. In this paper, we use a mathematical model that accounts for the interactions of primary factors regulating sleep timing and duration, namely the homeostatic sleep drive, the circadian rhythm and external light schedules, to simulate sleep behavior in 2.5 - 3 year old children and analyze causes of nighttime waking. Model results predict that children exhibiting earlier wake and sleep times may be more physiologically susceptible to nighttime waking. For such susceptible children, model results predict that variations in external light schedules, particularly delays in evening lights-off time, can induce nighttime waking. By simulating typical weekday vs weekend light schedules, model results provide practical guidance for regulation of external light schedules that can avoid sleep disruptions and promote consolidation of nighttime sleep.
\end{authorsummary}

\section{Introduction}
Sleep patterns in children show significant changes across development, from the irregular, polyphasic patterns of newborns, to more regular patterns with daytime naps for toddlers, and eventually to a single nighttime sleep episode similar to adult sleep patterns. Typically, by age 2 – 3 years old, young children settle into a sleep pattern with a single afternoon nap, a pattern which is maintained for the next $\sim 2$ years until around age 5 years old when the afternoon nap is dropped \cite{Davis2004}. During this period, while the qualitative sleep pattern is fairly consistent, young children can exhibit sleep issues such as bedtime problems, delayed sleep onset and nighttime waking \cite{Moore2007, Galland2010}.  Such sleep disturbances affect 20-30\% of young children \cite{Vriend} and can significantly disrupt not only the child’s well-being but also family functioning \cite{Meltzer2010}. Longitudinal data suggest that if untreated, these sleep problems may persist into later childhood, affecting cognitive, emotional, and behavioral development \cite{SleepMedicineReview2006Practice}. Among these various sleep concerns, nighttime waking has drawn increasing clinical and research attention due to its chronicity and impact on both the child and caregivers. 

Behavioral strategies, such as consistent bedtime routines and environmental modifications including the regulation of light exposure, have emerged as the most effective and durable treatment for nighttime waking in young children \cite{Galland2010, Davis2004, Meltzer2010}. External light is the primary regulator of the circadian rhythm, driven by the suprachiasmatic nucleus (SCN) which receives direct afferent projections from the retina. The circadian rhythm and its interaction with the homeostatic sleep drive, the unavoidable need for sleep after time spent awake, are well known to be salient determininants of sleep propensity and timing \cite{borbely}. Given the profound developmental implications of chronic sleep disruption and the behavioral treatment approaches, particularly those involving environmental regulation like light exposure, it is important to better understand how physiological and environmental factors combine to influence the occurrence of nighttime waking in young children.

In this paper, we use a mathematical model accounting for the interactions of the homeostatic sleep drive, the circadian rhythm and the effects of light on the circadian rhythm that is fit to experimental data on the sleep homeostat and sleep patterns in young children to analyze the effects of external light schedules on sleep patterns. The model is a recently developed extension of the classic Two-Process Model for the interactions of the homeostatic sleep drive, Process S, and the circadian rhythm, Process C \cite{Daan1984,tpm}. The new model, called the Homeostatic-Circadian-Light (HCL) model, includes a dynamic, van der Pol oscillator-based model for Process C that is fit to human arousal timing and can be entrained by external light input \cite{skeldon2023method}. We constrained the model to data for the dynamics of the homeostatic sleep drive measured in 2.5 – 3 year old children \cite{lebourgeois2012dynamics} and for typical napping sleep behavior in the same age group \cite{Akacem2015,athanasouli2024data} to find ranges of model parameter values that satisfy these constraints. Model solutions in these parameter ranges exhibited consolidated nighttime sleep, however, in some subsets of these ranges, model dynamics indicated that slight perturbations could induce a nighttime wake episode. While nighttime waking is often considered as a negative behavioral response to normal, spontaneous brief arousals during sleep \cite{Moore2007,Meltzer2010}, these model results suggest that there may be a physiological susceptibility for nighttime waking due to the interaction of sleep homeostasis and the circadian rhythm. 

To investigate how perturbations in external light schedules may influence the occurrence of nighttime waking, for a parameter set that is susceptible to nighttime waking, but does not exhibit nighttime waking for the typical external light schedule, we analyzed the effects of varying light schedules, including light intensity levels and timing of morning lights-on and evening lights-off times, on sleep patterns. Model results predict that changes in external light schedules can induce nighttime waking, with evening lights-off times having stronger effects. 

To provide predictions for practical behavioral strategies, we additionally simulated the effects of weekday vs weekend differences in light schedules that reflect typical schedules of working parents with children in childcare settings. Model results predict that delays in morning lights-on and evening lights-off times on the weekend can induce weekend nighttime waking, regardless of whether children take an afternoon nap on the weekend. Furthermore, weekend light changes were found to alter sleep timing during the week. This study highlights how sleep homeostasis, the circadian rhythm, external light, and parenting routines interact to shape sleep health in young children, providing a useful framework for improving sleep management practices.

\section{Results}

\subsection{Modeling typical sleep-wake behavior of young children}
\label{sec:default}

\begin{figure} [ht!]
\centering
\includegraphics[width=0.9\linewidth]{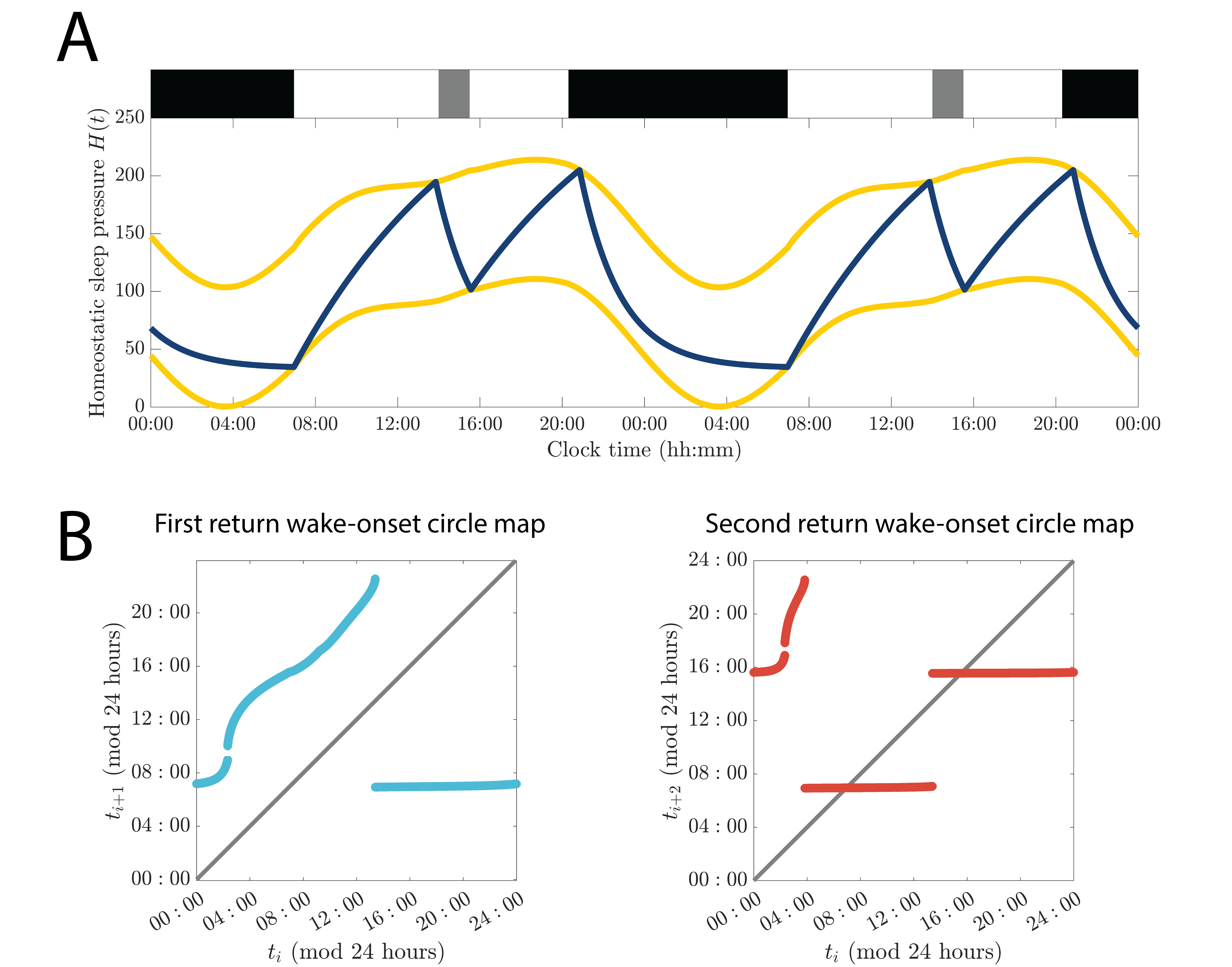}
\caption{\textbf{Time series and wake-onset circle maps of the HCL Model for the optimized parameter set.} (A) Simulated time series with the optimized parameter set showing wake, nap and sleep timing consistent with experimentally measured sleep-wake patterns in young children \cite{Akacem2015}. Dark ($I=0$), dim ($I=10$), and bright ($I=1000$) light conditions are indicated by black, gray, and white shading, respectively. (B) First and second return circle maps for wake onset times. The first return map plots wake-up time $t_{i+1}$ against the previous wake-up time $t_{i}$, while the second return map plots wake-up time $t_{i+2}$ against the second previous wake-up time $t_{i}$. The gray line represents the diagonal (identity line).}
\label{fig:1-optimized}
\end{figure}

We optimized the HCL model (see Model and Methods section) for a typical sleep pattern for 2.5 - 3 year old children \cite{Akacem2015, athanasouli2024data} restricting values for the homeostatic sleep drive parameters to ranges measured from sleep studies in this age group \cite{lebourgeois2012dynamics} (Table \ref{tab:data}, Fig \ref{fig:1-optimized}).  Optimization was achieved using a differential evolution algorithm \cite{buehren2025differential} to fit the homeostatic parameters ($\tau_{hw}$, $\tau_{hs}$, $h_{\max}$ and $h_{\min}$) and the amplitude and thresholds for the circadian process ($a$, $H_0^+$ and $H_0^-$), yielding the optimized parameter set in Table \ref{tab:optimized}. For the optimization, the external light schedule was set to match the experimental wake and bed times with a dim light period based on nap midpoint time \cite{athanasouli2024data}:  bright light (1000 lux) begins at 06:58, dim light (10 lux) starts at 14:00 and ends at 15:30, and lights are extinguished (0 lux) at 20:18. We call this the default light schedule. We note that the evening lights-off time is based on the reported bed time \cite{Akacem2015} which is earlier than the reported sleep onset time included in Table \ref{tab:data}. Simulated sleep metrics, including wake time, nap midpoint, nap duration, and sleep onset time, exactly matched the mean empirical observations listed in Table \ref{tab:data} (Fig \ref{fig:1-optimized}A).

\begin{table}[ht]
  \centering
  \begin{tabular}{|c|c||c|c|}
    \toprule
    \multicolumn{2}{|c||}{\textbf{Homeostatic sleep drive}} 
      & \multicolumn{2}{c|}{\textbf{Sleep metrics}}  \\
   \hline
  \textbf{Parameter} & \textbf{Value} 
  & \textbf{Metric} & \textbf{Value} \\
    \midrule
    $\tau_{hw}$ (day)                      
      & 0.3375 \,$\pm$\,0.0583                 
      & Wake time                
      & 06:58 \,$\pm$\, 00:30      \\
    $\tau_{hs}$ (day)                      
      & 0.0958 \,$\pm$\, 0.0208                  
      & Nap midpoint time        
      & 14:43 \,$\pm$\, 00:46      \\    
    $h_{\max}$ (\% SWA)                  
      & 314.5 \,$\pm$\, 51.7             
      & Nap duration (min)        
      & 102.6 \,$\pm$\, 20.1         \\
    $h_{\min}$ (\% SWA)                  
      & 26.7 \,$\pm$\, 6.9           
      & Sleep onset time         
      & 20:51 \,$\pm$\, 00:43      \\                                         
    \bottomrule
  \end{tabular}
  \caption{
 Experimentally observed ranges for homeostatic parameters \cite{lebourgeois2012dynamics} and sleep metrics  \cite{Akacem2015} in healthy children aged 2.5 to 3 years. Mean values and standard deviations are shown for the homeostatic rise and decay time constants ($\tau_{hw}$, $\tau_{hs}$), upper and lower bounds of homeostatic pressure ($h_{\max}$, $h_{\min}$), and five key sleep metrics: wake time, nap midpoint, nap duration, and sleep onset time.}
  \label{tab:data}
\end{table}

To survey ranges of parameter values that replicate the experimentally observed sleep patterns in young children, we varied the homeostat parameters ($\tau_{hw}$, $\tau_{hs}$, $h_{\max}$ and $h_{\min}$) across their experimental ranges and the circadian amplitude and baseline thresholds ($a$, $H_0^+$ and $H_0^-$) over large ranges and identified parameter sets that produced simulated sleep timings and durations within the experimentally observations \cite{Akacem2015} (Table \ref{tab:data}, see Model and Methods section). The feasible parameter ranges shown in Table \ref{tab:optimized} represent the smallest axis-aligned hypercube in the 7-dimensional parameter space that fully encloses the identified admissible parameter combinations. Notably, the feasible ranges for the homeostatic parameters closely match their empirically determined intervals. 

We note that not every parameter combination within the feasible ranges generates the experimentally observed sleep behavior. As an example, if we set \(H_0^+\) at its lower boundary while keeping all other parameters fixed at their optimized values, napping behavior is completely eliminated. However, simultaneous compensatory adjustments in parameters \(a\) and \(H_0^-\) within their feasible ranges recovers the empirical sleep patterns.

\begin{table}[ht]
  \centering
  \begin{tabular}{|c|c|c|c|}
    \toprule
  \textbf{Parameter} & \textbf{Optimized} & \textbf{Feasible range} & \textbf{Susceptible}\\
  & \textbf{values} & & \textbf{values} \\
    \midrule
    $\tau_{hw}$ (day)                      
      & 0.3849&  [0.2792 , 0.3958] & 0.3129\\
    $\tau_{hs}$ (day)                      
      & 0.0824&  [0.07502 , 0.1166] & 0.075\\    
    $h_{\max}$ (\% SWA)                  
      & 339.26&  [262.9 , 366.2] & 343\\
    $h_{\min}$ (\% SWA)                  
      & 33.6&  [19.81 , 33.6] & 32.5\\                                         
      $a$ (\% SWA)& 51.89&  [33.18 , 103.6] & 62\\
      $H_0^+$ (\% SWA)& 174.24&  [136.2 , 223.9] & 193\\
      $H_0^-$ (\% SWA)& 71.08&  [50.27 , 105] & 84.8\\
    \bottomrule
  \end{tabular}
  \caption{Parameter values for the HCL model that reproduce observed sleep metrics in healthy 2.5 to 3-year-old children listed in Table \ref{tab:data}. The homeostat parameters ($\tau_{hw}$, $\tau_{hs}$, $h_{\max}$, $h_{\min}$) are constrained to the experimentally measured ranges listed in Table \ref{tab:data}. Optimized values generate sleep-wake timing matching the mean sleep metrics in Table \ref{tab:data} and the feasible range represents the smallest axis-aligned hypercube that contains all sampled parameter sets producing sleep metrics within the observed ranges in Table \ref{tab:data}. Susceptible values generate sleep timing within the observed ranges but the model state is in close proximity to the bifurcation associated with nighttime waking. Remaining model parameters are set to values listed in Table \ref{tab:parameters}.}
  \label{tab:optimized}
\end{table}

To understand the stable periodic sleep-wake pattern, we consider the circle map for wake onset times, as has been done previously for the Two-Process Model \cite{bailey2018circle} and other sleep-wake regulation models \cite{BoothSIAMDSmap2017,AthanasouliSIAMDS2022,AthanasouliMathBiosci2023}. Fig \ref{fig:1-optimized}B shows the first (left panel) and second (right panel) return maps for the optimized parameter set. Here, we denote wake-up time as $t_i$, defined as when the homeostatic sleep pressure intersects with the lower circadian threshold, and the first and second subsequent wake-up time after $t_i$ as $t_{i+1}$ and $t_{i+2}$, respectively. As is the case for the Two-Process Model, the maps exhibit discontinuities or gaps due to the piecewise-smooth homeostatic sleep drive \cite{bailey2018circle}. The first return map does not intersect with the $y=x$ diagonal indicating that, in the stable periodic solution, sequential wake onsets do not occur at the same time of day. This is expected since the next wake onset after morning wake-up time is the nap wake-up time.  
However, the second return map shows two intersection points with the $y=x$ diagonal representing stable fixed points, one at the morning wake onset time and the other at the nap wake-up time. Dynamically, these fixed points indicate that over time, regardless of initial conditions, the sleep-wake pattern converges to consistent daily morning and nap wake-up times.

While the model exhibits consolidated nighttime sleep with this optimized parameter set, the potential for night waking is suggested by how close the sleep homeostat trajectory comes to the lower circadian threshold curve during the nighttime sleep episode (around midnight). Specifically, if model parameters were varied such that the homeostat trajectory made a tangent intersection with the lower circadian threshold curve near that time, the model would predict a nighttime waking. 
This bifurcation in the model solution would be reflected in a change in the second return circle map such that the map curves would no longer intersect the $y=x$ diagonal and a border-collision bifurcation would have occurred \cite{bailey2018circle}.

To explore how differences in physiological parameters affect the proximity of toddlers' sleep patterns to this bifurcation associated with nighttime waking, we systematically investigated effects of parameter variation within the identified feasible ranges. 
We surveyed model solutions as each of these seven parameters were individually varied toward their respective boundaries within the ranges listed in Table \ref{tab:optimized}. This allowed us to qualitatively assess how each parameter influences the model’s proximity to the bifurcation. Table \ref{tab:bifurcation} summarizes the direction of each parameter’s variation that moves the model closer to the bifurcation as well as its physiological significance and effect. 
Our qualitative analysis shows that faster build-up or dissipation of the homeostatic sleep drive shifts the model towards the bifurcation. Likewise, higher maximum and lower minimum asymptotes for the homeostat have a similar effect, which can be expected because these variations influence the rates of change of the exponential homeostatic drive.
In terms of inducing the bifurcation in the model, decreasing $\tau_{hs}$ or $h_{\min}$, as well as increasing $h_{\max}$, push the system closer to but do not induce the bifurcation. However, decreasing $\tau_{hw}$ directly induces the bifurcation. Notably, these parameter variations that shift the model closer to the bifurcation produce earlier sleep phenotypes. Furthermore, a smaller gap between circadian thresholds (smaller $H_0^+$ and larger $H_0^-$) similarly moves the system toward the bifurcation and promotes earlier sleep timing. These results suggest that children exhibiting earlier sleep phenotypes may be physiologically vulnerable to nighttime waking due to interactions of their homeostatic sleep drive and circadian rhythm.

We note that because the model operates within a high-dimensional parameter space, several parameters ($h_{\max}$, $h_{\min}$, $a$, $H_0^+$, and $H_0^-$) compensate for each other, allowing multiple parameter combinations to yield similar sleep metrics and place the model near the bifurcation.

\begin{table}
  \centering
  \renewcommand{\arraystretch}{1.3}
  \begin{tabular}{|c|p{2.5 cm}|p{8.5 cm}|}
    \hline
    \textbf{Parameter} & \textbf{Shift towards} & \textbf{Physiological Mechanism \& Sleep phenotype} \\
    & \textbf{bifurcation} & \\
    \hline
    $\tau_{hw}$ & Decrease & Smaller $\tau_{hw}$ increases rate of build-up of sleep pressure during waking and induces earlier sleep timing   \\
    \hline
    $\tau_{hs}$ & Decrease  & Smaller $\tau_{hs}$ leads to faster dissipation of sleep pressure during sleep and induces earlier sleep timing   \\
    \hline
    $h_{\max}$  & Increase  & Higher $h_{\max}$ increases the maximum sleep pressure, consequently increasing sleep pressure build-up rate during waking, and induces earlier sleep timing   \\
    \hline
    $h_{\min}$  & Decrease & Lower $h_{\min}$ reduces baseline sleep pressure, consequently increasing dissipation of sleep pressure during sleep, and induces earlier sleep timing  \\
    \hline
    $a$ & Decrease  & Smaller $a$ decreases circadian amplitude thus reducing circadian influence on sleep timing   \\
    \hline
    $H_0^+$ & Decrease & Smaller $H_0^+$ reduces sleep pressure level at sleep onset leading to earlier sleep onset timing   \\
    \hline
    $H_0^-$ &  Increase & Larger $H_0^-$ increases sleep pressure level at wake onset resulting in earlier wake onset timing   \\
    \hline
  \end{tabular}
  \caption{Summary of parameter variations that shift the model closer towards the bifurcation associated with nighttime waking, the physiological mechanism represented by the variation and the resulting change in sleep phenotype.}
  \label{tab:bifurcation}
\end{table}

\begin{figure}
\centering
\includegraphics[width=0.9\linewidth]{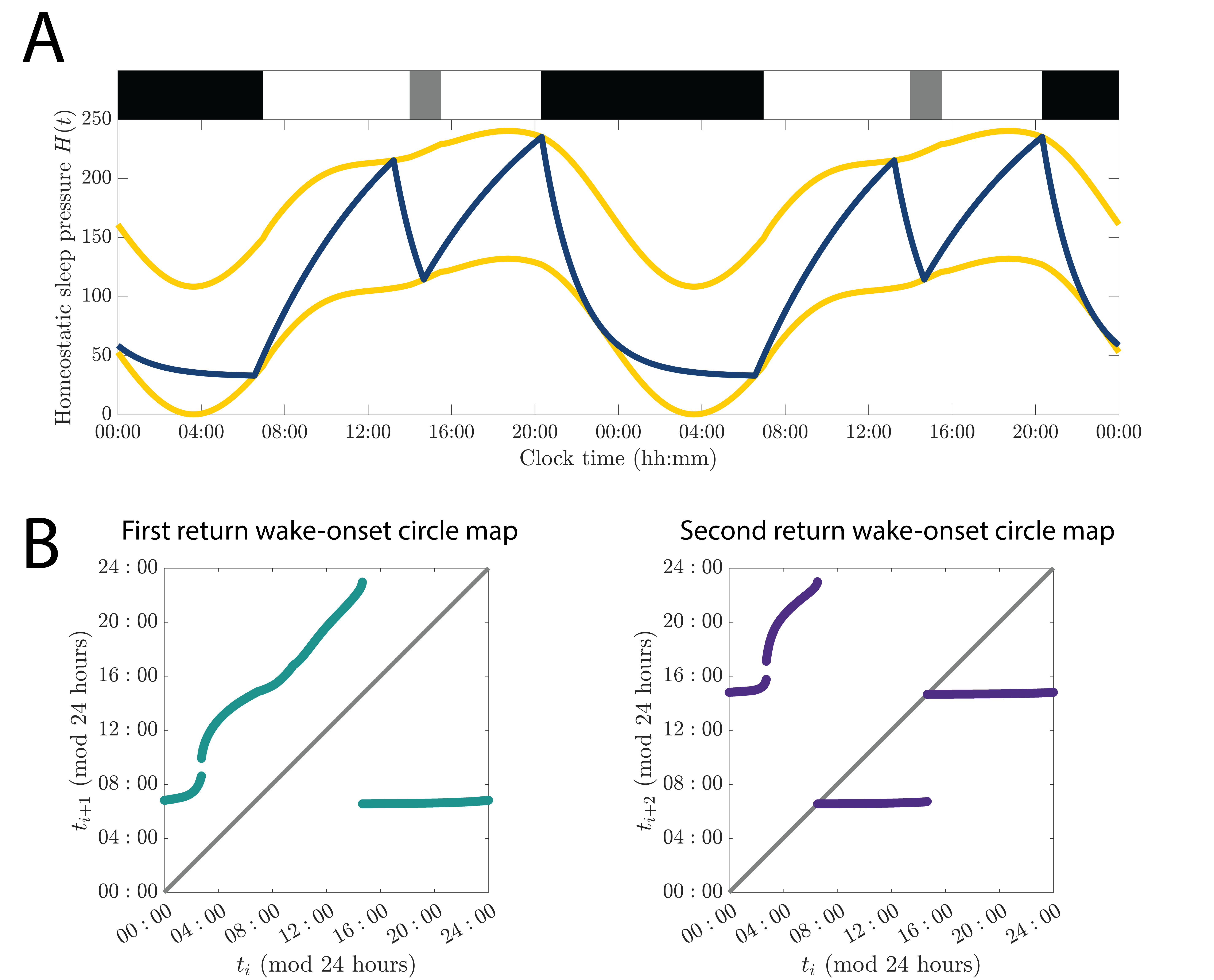}
\caption{\textbf{Time series and wake-onset circle maps of the HCL model for the susceptible parameter set.} Simulation conditions and layout are identical to Figure~\ref{fig:1-optimized}. Resolution of homeostat and circadian trajectory curves gives the appearance of a tangent intersection but it does not occur. }
\label{fig:1-default}
\end{figure}

We further explored simultaneous variations in all seven parameters, identifying an alternate parameter set near the bifurcation, which we denote as the susceptible parameter set as it is susceptible to nighttime waking (Table \ref{tab:optimized}). 
Guided by the qualitative rules summarized in Table \ref{tab:bifurcation}, this susceptible parameter set aligns closely with the relative changes suggested for $\tau_{hw}$, $\tau_{hs}$, $h_{\min}$, and $h_{\max}$ compared to the optimized parameter set, but has slight compensatory offsets in parameters $a$, $H_0^+$, and $H_0^-$ .  For the susceptible parameter set with the default external light schedule, model simulations display a morning wake-up time at 6:33, nap initiation at 13:13, nap conclusion at 14:40, and nighttime sleep onset at 20:19 (Fig \ref{fig:1-default}A). The model time trace clearly shows the proximity of the homeostat trajectory to the lower circadian threshold during the nighttime sleep episode (there isn't a tangent intersection between the curves despite the appearance of one due to figure resolution). We also calculated the first and second return circle maps for wake onset (Fig \ref{fig:1-default}B). Similar to the optimized parameter set, the first return map shows no intersection with the $y=x$ diagonal line, but the second return map displays two intersection points, representing stable morning and nap wake-up times. The proximity of the fixed points to the ends of the map curves confirms that with these parameter values the model is indeed near the border-collision bifurcation associated with nighttime waking. 

Considering this susceptible parameter set as representative of sleep patterns of young children susceptible to nighttime wakings, in the following sections we analyze how changes in external light schedules can influence sleep timing and induce nighttime waking. In the HCL model, changes in the external light schedule introduce slight variations in the circadian threshold curves, particularly at light onset and offset times. Since in the susceptible parameter set the homeostat trajectory and the lower circadian threshold curve pass very close to each other, the influence of such slight variations, especially on nighttime sleep onset timing, may cause them to intersect and create a nighttime wake episode. As our results show, variations in light intensity levels and timing of morning lights-on or evening lights-off times can induce this nighttime waking. We also analyze how changes in weekend external light schedules can disrupt sleep behavior and influence sleep timing during the week.



\subsection{Low daytime lux levels can promote nighttime waking}
\label{sec:intensity}

\begin{figure}[h!]
    \centering
    \includegraphics[width=0.8\linewidth]{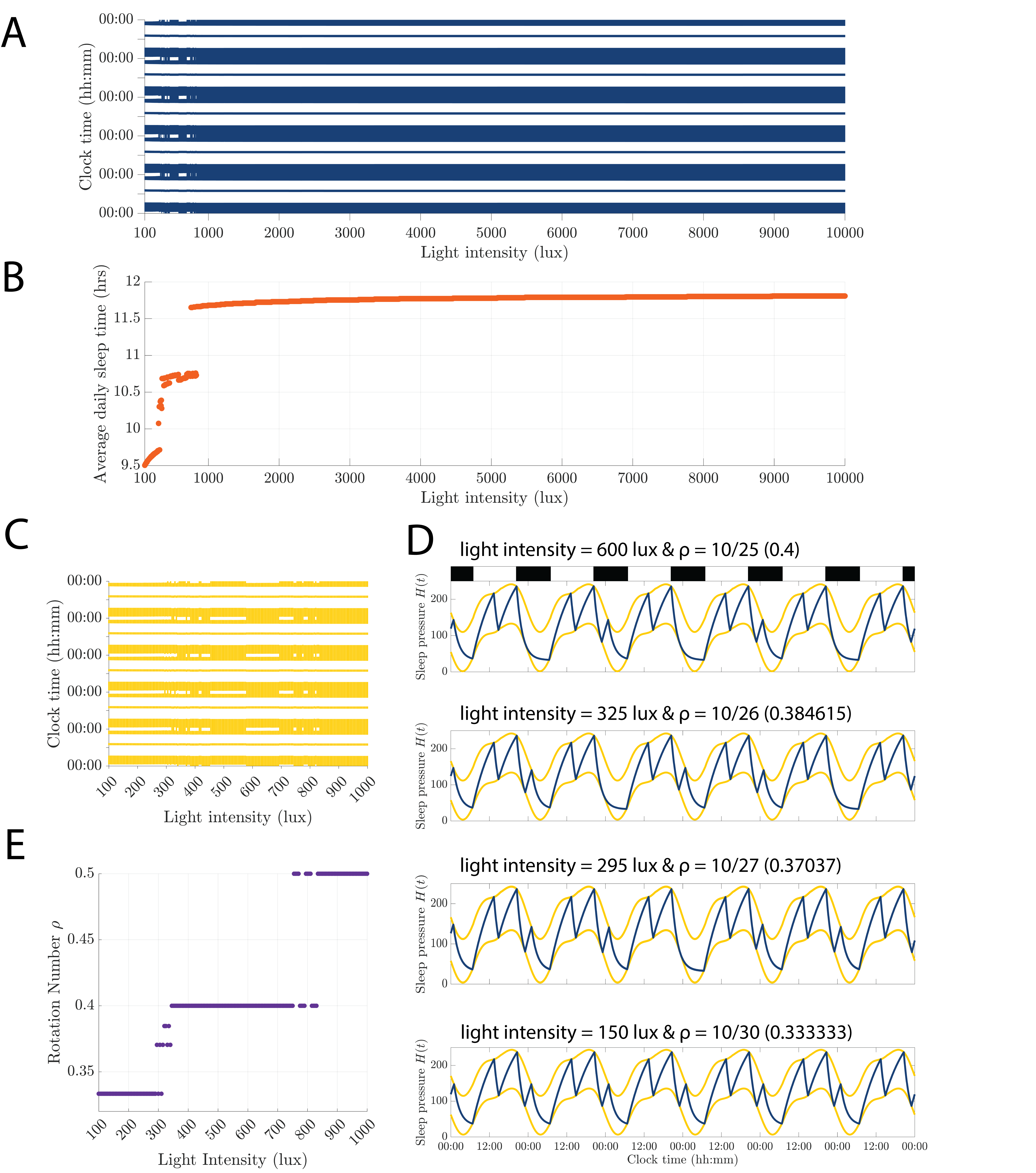}
    \caption{\textbf{Nighttime waking is induced by low daytime light intensity.} (A) Simulated sleep patterns over a five-day interval (y-axis, blue intervals indicate when sleep occurred) under daytime light intensities ranging from 100 lux to 10,000 lux (x-axis) for the default external light schedule. (B) Average daily sleep duration as a function of daytime light intensity. (C) Detailed visualization of sleep phenotypes at lower daytime light intensities (100–1,000 lux), illustrating specific nighttime waking patterns. (D) Representative time series illustrating distinct sleep phenotypes over 6 consecutive days under specific light intensities (top to bottom panels: 600 lux, 325 lux, 295 lux, and 150 lux). (E) Bifurcation diagram showing stable sleep solutions, quantified by the rotation number $\rho$ (number of days divided by number of sleep episodes), across daytime intensities from 100 to 1,000 lux. Rotation numbers are listed in the examples in (D). }
    \label{fig:2}
\end{figure}

In this section, we explore the role of daytime light intensity in shaping sleep patterns. We simulated the HCL model with the susceptible parameter set and the default external light schedule for daytime light intensities ranging from 100 lux to 10,000 lux (light intensity during the nap remained at 10 lux, Fig \ref{fig:2}A). 
We found that light intensities higher than the default value of 1000 lux caused a slight increase in the average daily sleep duration (Fig \ref{fig:2}B), however lower lux values introduced nighttime sleep disruptions, specifically a period of nighttime waking with a duration of approximately 2 hours occurring roughly between 21:00 and 2:00. At the lowest daytime light intensities ($\sim 100-300$ lux), nighttime wakings occurred every night, but they occurred less frequently for higher lux values. 

To investigate phenotypic variations at lower daytime light intensities, we analyzed sleep dynamics in greater detail within the 100 to 1,000 lux range (Fig \ref{fig:2}C). The highest intensity associated with recurrent nighttime waking was approximately 830 lux, characterized by nighttime wakings occurring every other day. To systematically quantify sleep disruption phenotypes, we introduced the rotation number $\rho$, defined as the ratio of circadian days to sleep episodes in stable sleep patterns. Using data over 10 circadian days, we plotted $\rho$ against light intensities ranging from 100 to 1,000 lux and identified five distinct rotation numbers corresponding to unique disruption phenotypes (Fig \ref{fig:2}E). The non-monotonic relationship between rotation number and light intensity, specifically for lux levels between 750 - 850, shows the inherent complexity of the HCL model governing toddler sleep patterns.

We further show these phenotypes by presenting representative time traces of sleep dynamics at different daytime light intensities (Fig \ref{fig:2}D). With decreasing intensity, the observed phenotypes transitioned as follows: (1st row) sleep disruptions every other day (600 lux, $\rho=10/25$), (2nd row) two disrupted days followed by one undisturbed day (325 lux, $\rho=10/26$), (3rd row) three disrupted days followed by one undisturbed day (295 lux, $\rho=10/27$), and ultimately (4th row) daily sleep disruptions (150 lux, $\rho=10/30$).

\subsection{Longer daytime bright-light exposure induces nighttime waking}
\label{sec:timing}

We examined how different durations of daytime bright light influence sleep patterns, specifically investigating the effects of shifting the morning onset and evening offset timing of bright-light exposure. To explore how changing the morning lights-on time affects sleep timing, we defined the lights-on shift as $\Delta t_{on}$, comparing it to the default schedule of 6:58, where positive $\Delta t_{on}$ values represent shifts to later times and negative values indicate shifts to earlier times. Similarly, we defined $\Delta t_{off}$ to denote shifts in the evening lights-off time compared to the default schedule of 20:18, following the same sign conventions as $\Delta t_{on}$. Additionally, we explored the combined effect of extending or shortening bright-light exposure durations by symmetrically shifting both morning lights-on and evening lights-off times. In this case, $\Delta t$ signifies the magnitude of simultaneous shifts of morning lights-on and evening lights-off times — positive values indicate extended bright-light durations, and negative values indicate shorter durations.

\begin{figure} [ht!]
\centering
\includegraphics[width=\linewidth]{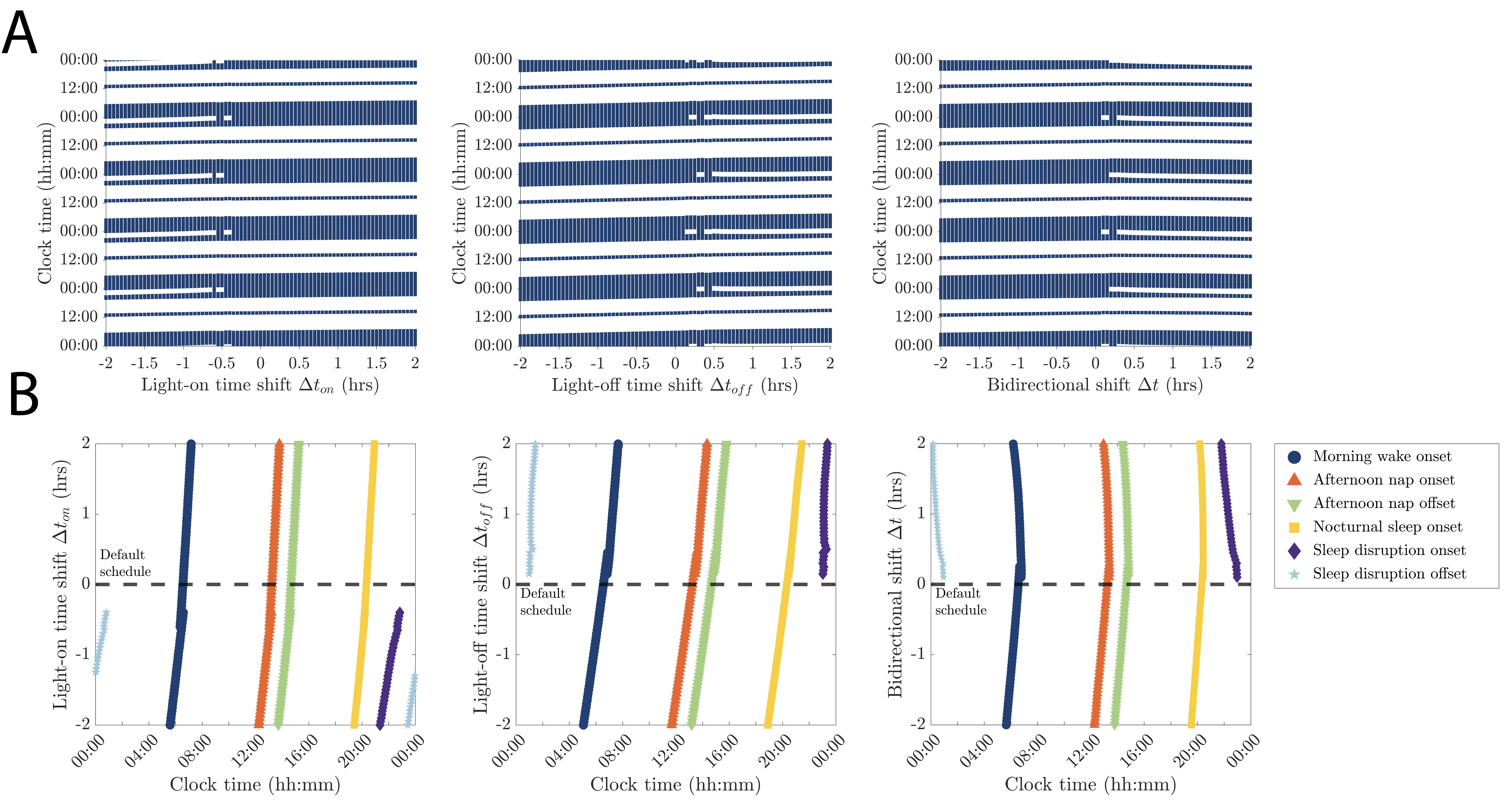}
\caption{\textbf{Longer daytime bright-light exposure promotes nighttime waking.} (A) Simulated sleep patterns over a five-day interval (y-axis, blue intervals indicate when sleep occurred) for different schedules of daytime bright-light exposure. Left panel: shift in morning lights-on time by $\Delta t_{on}$ (h); Middle: shift in evening lights-off time by $\Delta t_{off}$ (h); Right: simultaneous shift in morning lights-on and evening lights-off time, each by $\Delta t$ (h) (negative (positive) values indicate shifts to earlier (later) times). Other light timings were set at the default external light schedule.(B) Sleep and wake onset times across the 24 h day (x-axis) for the different daytime bright-light exposure schedules (y-axis): dark blue = morning wake-up time, orange = nap onset time, green = nap wake-up time, yellow = evening sleep onset time, purple = nighttime waking onset time, light blue = nighttime waking offset time. Panels show the same bright-light variations as in A. Where curves show two values for the same light shift indicates an alternating pattern of nighttime waking and consolidated nighttime sleep.}
\label{fig:3}
\end{figure}

Simulated sleep patterns across five days for each scenario are shown in Fig \ref{fig:3}A. The simulations revealed that longer daytime bright-light durations generally led to sleep disruptions, whether this occurred through earlier morning lights-on times (left panels) or later evening lights-off times (middle panels) or simultaneous shifts (right panels). Sensitivity to the direction of extension in bright-light duration was not symmetric. In particular, morning lights-on time needed to be earlier than 6:34 ($\Delta t_{on} \leq -0.4$) to cause nighttime waking. However, even a slight delay in evening lights-off time past 20:27 ($\Delta t_{off} \geq 0.15$) generated nighttime waking. Interestingly, reducing the duration of bright-light exposure by any means did not induce sleep disruptions.

Next, we examined how these varying light schedules impacted sleep and wake timing. We identified the changes in morning wake onset, afternoon nap onset, afternoon nap offset, evening sleep onset, and the onset and offset of nighttime waking in the stable sleep patterns (Fig \ref{fig:3}B). Generally, shifting either morning lights-on time or evening lights-off time earlier led to earlier times for all sleep-wake transitions, and shifting them later delayed all transitions. However, the shifts in evening lights-off time had greater effects on transition times. 
For instance, turning lights off two hours earlier resulted in toddlers waking up 1 hour and 30 minutes earlier, whereas turning lights on two hours earlier leads to waking up about 58 minutes earlier. This result suggests a greater sensitivity to changes in evening lights-off time compared to morning lights-on time in young children.
Additionally, simultaneous adjustments to both morning lights-on and evening lights-off times generally shifted all events earlier. 

These insights offer practical guidance for altering sleep-wake patterns without inducing nighttime waking. For example, to promote a later morning wake time without disrupting nighttime sleep, the most effective strategy is to turn lights on later in the morning rather than turning them off later in the evening. Conversely, to encourage earlier morning wake times or naps, turning the lights off earlier in the evening is less disruptive than turning lights on earlier in the morning. 


\subsection{Effects of realistic weekly external light schedules on sleep patterns}
\label{sec:weekly-schedule}

In the previous sections, we examined how consistent daily light schedules impact stable sleep patterns. Here, we extend our analysis to realistic weekly variations in light schedules, inspired by typical differences in parental activities between weekdays and weekends. Since parents typically work during weekdays following a regular routine but follow varied schedules on weekends, we defined a baseline weekday schedule identical to the default external light schedule detailed in Section \ref{sec:default}. Specifically, the weekday (Monday thru Friday) morning lights-on time is 6:58, and evening lights-off time is 20:18, with dim light during the afternoon naptime (from 14:00 to 15:30). For the weekends (Saturday and Sunday), we assumed continuous daytime bright light (1000 lux) from morning lights-on time to evening lights-off time without afternoon dim-light periods, reflecting conditions where parents may not actively manage afternoon nap opportunities or lighting. We assume this would correspond to the child sleeping in their car seat or stroller, or in a room without dimmed light levels.

\begin{figure}
\centering
\includegraphics[width=\linewidth]{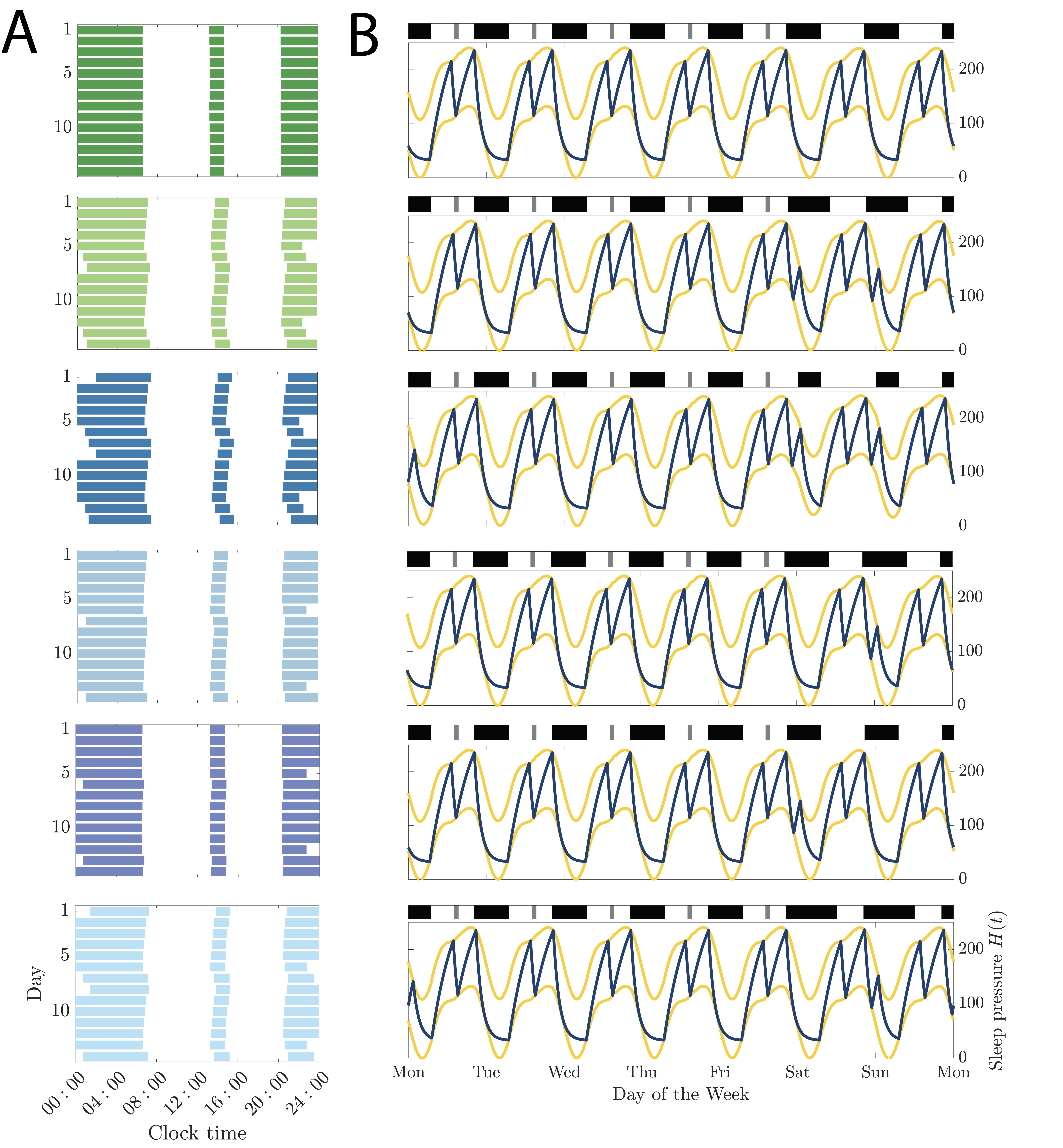}
\caption{\textbf{Examples of possible weekly sleep patterns obtained under varying weekend external light schedules.} (A) Daily sleep timing (x-axis, colored bars indicate sleep times) across 14 days (y-axis, day 1 = Monday) for the six representative scenarios. (B) Corresponding model simulation time traces depicting homeostatic sleep pressure (blue) and circadian thresholds (maize) with associated external light schedule (top). For (weekend morning lights-on time, weekend evening lights-off time): (first row): (6:58, 20:18) no nighttime wakings, (2nd row): (9:00, 22:00) Friday and Saturday nighttime wakings, (3rd row): (7:12, 24:00) Friday, Saturday, and Sunday nighttime wakings, (4th row): (10:00, 20:18) Saturday only nighttime waking, (5th row): (6:58, 20:30) Friday only nighttime waking, (bottom row): (12:00, 20:18) Saturday and Sunday nighttime waking.}
\label{fig:4}
\end{figure}

\begin{figure}[h]
\centering
\includegraphics[width=0.7\linewidth]{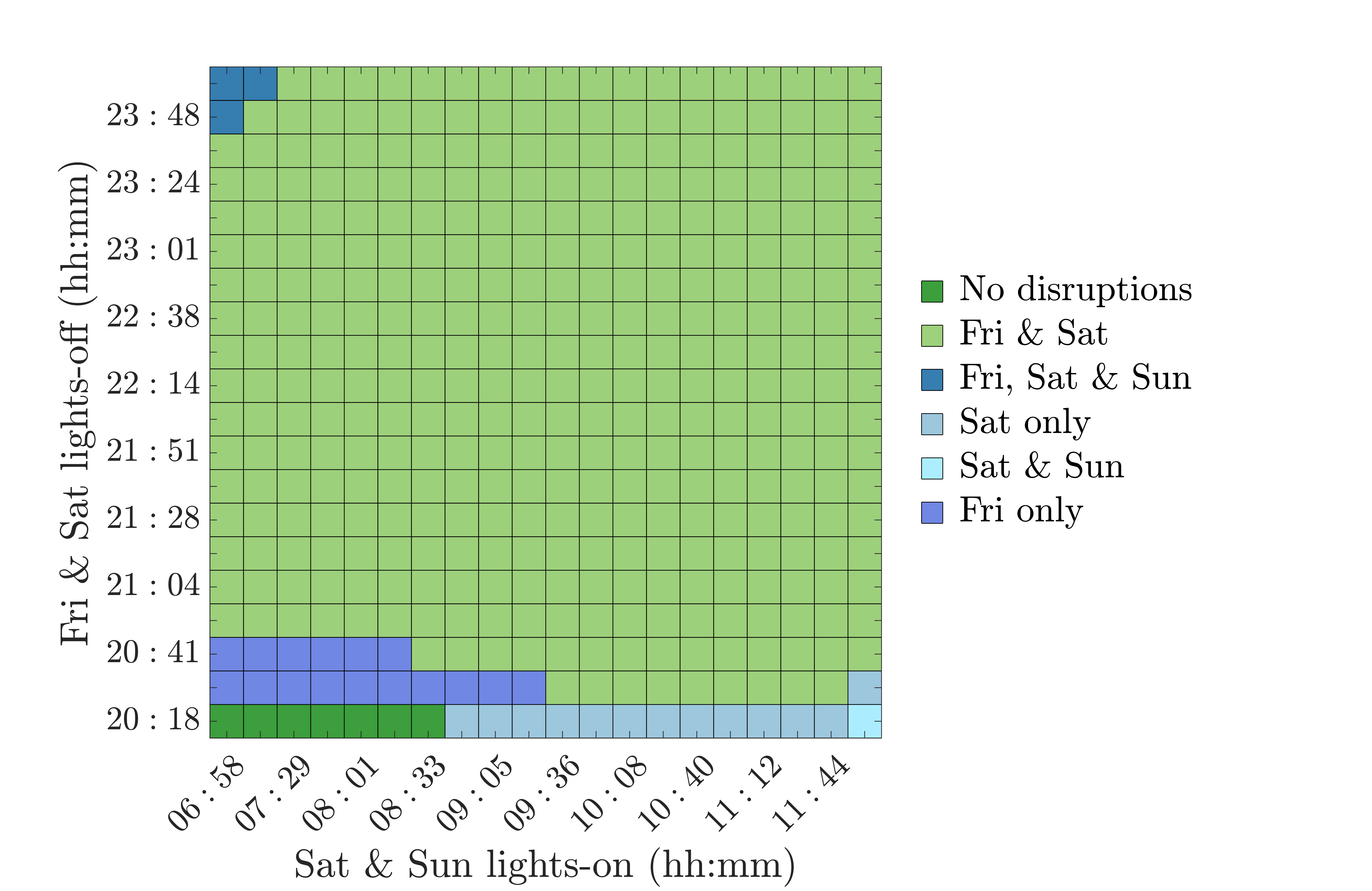}
\caption{\textbf{Occurrence of nighttime sleep disruptions due to varying delays in weekend morning lights-on and evening lights-off times.} Predicted sleep patterns resulting from varied delays in Friday and Saturday lights-off time (y-axis) and Saturday and Sunday morning lights-on time (x-axis). Colors correspond to the example sleep patterns shown in Fig \ref{fig:4}: no nighttime wakings (dark green); Friday and Saturday nighttime wakings (light green); Friday, Saturday and Sunday nighttime wakings (blue); Saturday only nighttime waking (blue-gray); Friday only nighttime waking (violet-blue); Saturday and Sunday nighttime waking (light blue). Grid spacing is $\sim 16$ minutes on the x-axis and $\sim 12$ minutes on the y-axis.}
\label{fig:5}
\end{figure}

To simulate realistic weekend variations, we introduced two adjustable parameters: a delayed morning lights-on time $t_{on}$ (reflecting parents waking later on weekends) and a delayed nighttime lights-off time $t_{off}$ (representing a delayed bedtime). These adjustments occur on Saturday and Sunday mornings, and Friday and Saturday evenings, respectively. We assumed that evening lights-off time on Sunday was the same as during the week, consistent with the need to resume regular weekly routines. 
We simulated all combinations of $(t_{on}, t_{off})$, with $t_{on}$ ranging from 6:58 to 12:00 and $t_{off}$ ranging from 20:18 to midnight. Our simulations identified six distinct weekly sleep patterns (Fig \ref{fig:4}) that occurred for different weekend external light schedules (Fig \ref{fig:5}).

We found that maintaining the default weekday light schedule without the afternoon dim-light periods on weekends did not cause nighttime sleep disruptions or alter sleep timing from the weekday schedule (Fig \ref{fig:4}, first row; Fig \ref{fig:5}, dark green). And the weekday schedule was preserved with up to a 1.5 h delay in weekend morning lights-on time without changing the evening lights-off time. However, shifting the morning lights-on time later beyond approximately 1.5 hours on weekends generated nighttime sleep disruptions on Saturday nights and subsequent later shifts in Sunday’s sleep timing (Fig \ref{fig:4}, 4th row; Fig \ref{fig:5}, blue-gray). Sleep-wake times gradually returned to baseline timing during the following weekdays. Extremely late morning lights-on times (around noon) induced nighttime wakings on both Saturday and Sunday nights (Fig \ref{fig:4}, bottom row; Fig \ref{fig:5}, light blue).

Delaying only the weekend evening lights-off time resulted in immediate and pronounced disruptions due to greater sensitivity to delayed lights-off as reported above. Even minor delays in Friday night lights-off time initiated nighttime waking on Friday night (Fig \ref{fig:4}, 5th row; Fig \ref{fig:5}, violet-blue), and further delays extended disruptions across both Friday and Saturday nights (Fig \ref{fig:4}, 2nd row; Fig \ref{fig:5}, light green). Ultimately, nighttime waking occurred on Friday, Saturday and Sunday nights if the delay in lights-off time reached around midnight (Fig \ref{fig:4}, 3rd row; Fig \ref{fig:5}, blue). Again, the weekday default schedule gradually restored sleep timings to baseline conditions during the following weekdays.

In summary, delays in weekend evening lights-off time, reflecting later weekend bedtimes, induced weekend nighttime waking and simultaneous delays in Friday and Saturday evening lights-off, and Saturday and Sunday morning lights-on times usually caused nighttime wakings on both Friday and Saturday nights. 

\subsection{Impact of weekend activities eliminating toddler naps on weekly sleep patterns}
\label{sec:no-nap}

In the previous section, we examined how weekend variations in bright-light schedules influence sleep patterns, assuming sleep was permitted in the afternoon for a nap but the external light level was not changed.  However, in realistic situations, parents frequently engage in weekend activities or outings that would not be conducive for an afternoon sleep opportunity. During such outings or activities, children may remain awake despite accumulating high sleep pressure exceeding their circadian upper threshold. This enforced afternoon wakefulness can impact subsequent sleep timing and duration.

To explore this scenario, we modeled conditions where sleep was prevented during Saturday and Sunday afternoons, and sleep was permitted only after evening lights-off on those days. Due to high sleep pressure, the model predicted sleep onset immediately upon evening lights-off. As above, we simulated all combinations of Saturday and Sunday morning lights-on times, and Friday and Saturday evening lights-off times, ranging from 6:58 to 12:00 for morning lights-on and 20:18 to midnight for lights-off. On Sunday night, the weekday light schedule was resumed with evening lights-off at 20:18. 

\begin{figure}[h]
\centering
\includegraphics[width=\linewidth]{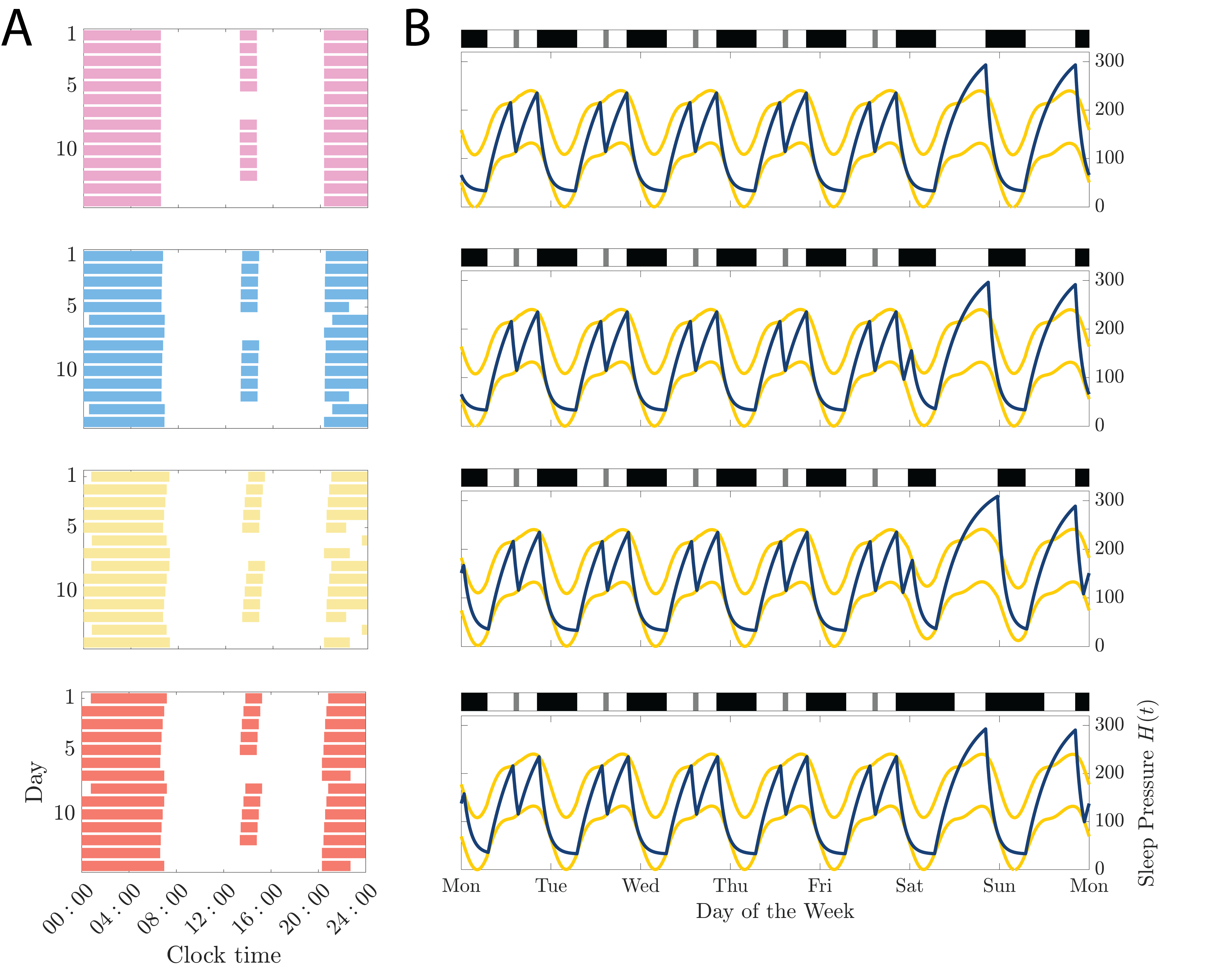}
\caption{\textbf{Effect of preventing afternoon naps on sleep patterns under varying delays in weekend morning lights-on and evening lights-off times.} (A, B) Example sleep patterns obtained with different weekend light schedules. Format is the same as in Fig \ref{fig:4}A and B. For (weekend morning lights-on time, weekend evening lights-off time): first row: (6:58, 20:18) no nighttime wakings , 2nd row: (6:58, 21:00) Friday only nighttime waking, 3rd row: (6:58, 23:30) Friday and Sunday nighttime waking, 4th row: (12:00, 20:18) Sunday only nigttime waking. }
\label{fig:6}
\end{figure}

\begin{figure}[h]
\centering
\includegraphics[width=0.7\linewidth]{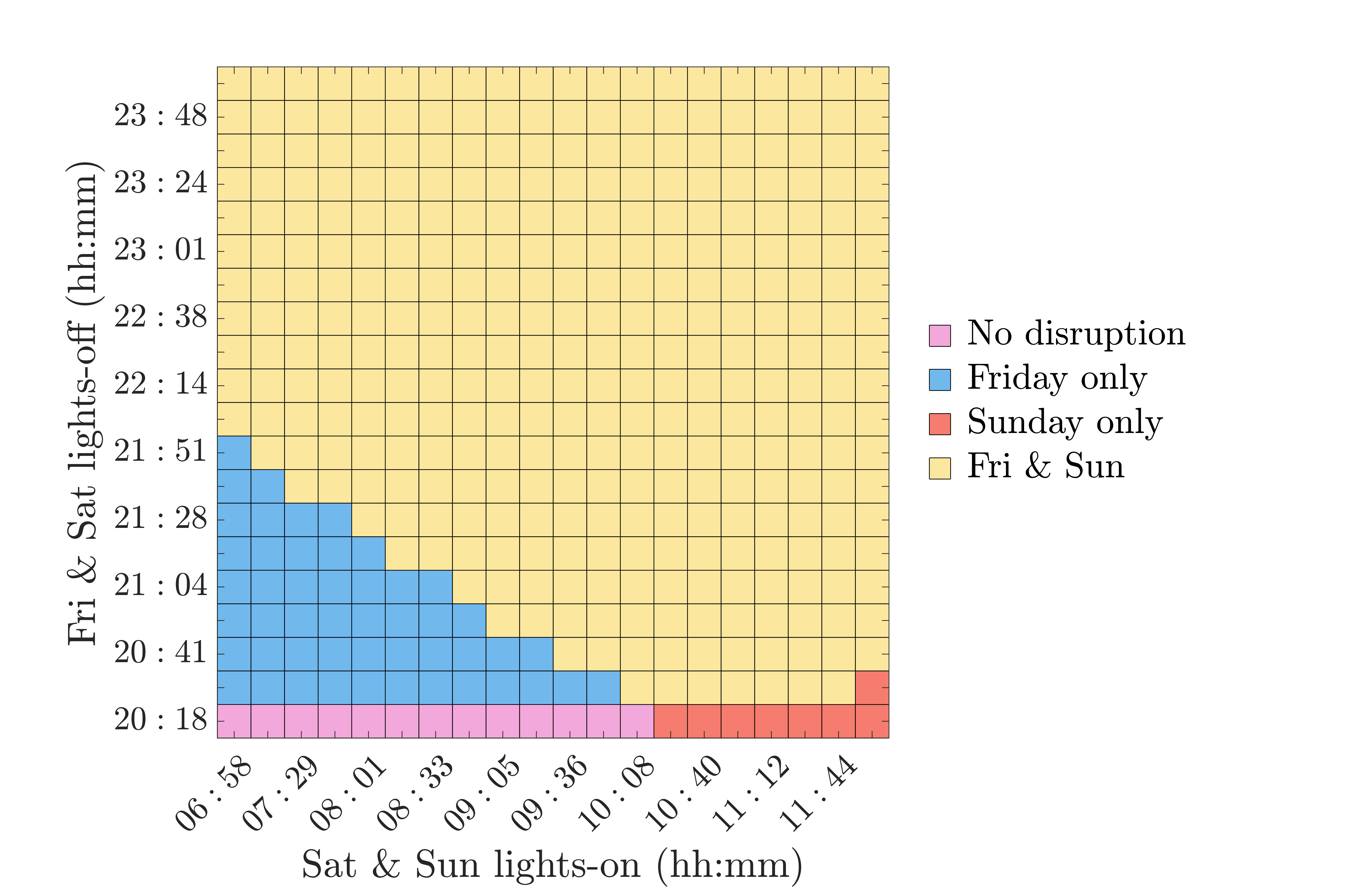}
\caption{\textbf{Occurrence of nighttime sleep disruptions due to varying delays in weekend morning lights-on and evening lights-off times when afternoon naps are prevented.} Predicted sleep patterns resulting from varied delays in Friday and Saturday lights-off time (y-axis) and Saturday and Sunday morning lights-on time (x-axis). The prevention of sleep during Saturday and Sunday afternoons causes the homeostatic sleep drive to increase beyond the upper circadian threshold and sleep onset to occur at evening lights-off time. Colors correspond to the example sleep patterns in Fig \ref{fig:6}A, B: no nighttime wakings (pink); Friday only nighttime waking (blue); Sunday only nighttime waking (orange); Friday and Sunday nighttime waking (yellow). Grid spacing is $\sim 16$ minutes on the x-axis and $\sim 12$ minutes on the y-axis.}
\label{fig:7}
\end{figure}

Our results identified four distinct sleep patterns (Fig \ref{fig:6}) that occurred for different weekend light schedules (Fig \ref{fig:7}). Fixing Friday and Saturday nights' lights-off time to the weekday schedule (20:18) while progressively delaying weekend morning lights-on maintained consolidated nighttime sleep (Fig \ref{fig:6}, 1st row; Fig \ref{fig:7}, pink), except that excessive delays ($> 3$ h) induced nighttime waking on Sunday night only (Fig \ref{fig:6}, 4th row; Fig \ref{fig:7}, orange).

If Friday and Saturday evening lights-off times were delayed, a region of parameter space emerged where nighttime waking occurred only on Friday night (Fig \ref{fig:6}, 2nd row; Fig \ref{fig:7}, blue). With longer delays in Saturday and Sunday lights-on time, both Friday and Sunday nighttime waking was induced (Fig \ref{fig:6}, 3rd row; Fig \ref{fig:7}, yellow). With long enough delays in weekend lights-on and lights-off times, this pattern of nighttime waking on Friday and Sunday nights was consistently exhibited. 

The primary effect of eliminating weekend afternoon sleep was the consolidation of sleep on Saturday night. The higher homeostatic sleep drive on Saturday modulated its timing relative to the circadian rhythm to prevent a nighttime wake episode. Interestingly, for moderate delays in weekend light timing, nighttime waking was predicted only on Friday night due to later evening lights-off time, suggesting that higher sleep pressure promoted consolidated nighttime sleep on both Saturday and Sunday nights. However, longer delays in morning lights-on and/or evening lights-off times introduced a nighttime wake episode on Sunday night, in addition to Friday night.



Notably, all patterns of weekend nighttime waking introduced shifts in Monday sleep-wake transitions to later times (Fig \ref{fig:6}A, bottom 3 rows). Sleep timings then gradually shifted earlier over the following weekdays indicating a prolonged effect of the weekend disruptions.

\section{Discussion}

Nighttime waking in toddlers and pre-school age children is a common sleep problem and is generally considered to be caused by negative attitudes towards sleep or the inability of the child to return to sleep without intervention by parents or caregivers, such as rocking or lying down with the child \cite{Moore2007}. Our model results suggest a potential physiological susceptibility for nighttime waking in children 2.5 – 3 years old due to interactions of the homeostatic sleep drive and the circadian rhythm. Notably, parameter sets associated with earlier sleep phenotypes (small $\tau_{hw}$, $\tau_{hs}$, $h_{\min}$, and large $h_{\max}$) appear inherently more susceptible to nighttime waking.

Our results further demonstrate that variations in external light schedules can promote nighttime waking when there is this susceptibility. Specifically, our results indicate that delays in evening lights-off time and, generally, extensions in bright-light exposure duration are more likely to induce nighttime waking. For example, 30 minute delays in evening lights-off time may be sufficient to induce a nighttime wake episode. Shortening bright-light duration by delaying morning lights-on time, shifting evening lights-off time earlier or both simultaneously by up to 2 h did not cause nighttime sleep disruptions. Interestingly, for weekly variations in external light schedules consisting of delays in morning lights-on and evening lights-off times on the weekend, most delays caused nighttime waking on at least one of the weekend nights, with nighttime waking on two weekend nights occurring for most combinations of delays. Nighttime sleep disruptions occurred even when afternoon naps were skipped, and the homeostatic sleep drive was higher than normal at evening bedtime. These weekend sleep disruptions were observed to cause delays in Monday sleep timing, especially when weekend naps were skipped, that slowly shifted earlier throughout the week. These findings suggest the complex trade-offs parents face when trying to strategically time weekend activities without inducing sleep disruption that may have prolonged effects during the week.  

To model sleep behavior in young children, we used experimentally determined ranges for parameters for the homeostatic sleep drive \cite{lebourgeois2012dynamics} and for typical sleep behavior in 2.5 – 3 year olds \cite{Akacem2015,athanasouli2024data}.  And we optimized values of parameters that affect the amplitude of the circadian rhythm ($a$) and the baseline thresholds for the circadian processes ($H_0^+$ and $H_0^-$) that govern the tolerance to increases and decreases, respectively, in sleep pressure for causing a sleep-wake transition. We assumed that the daily profile of the circadian propensity for wakefulness was the same as identified for adults in the HCL model \cite{skeldon2023method}. There is evidence that circadian rhythms are weak during the first few weeks after birth \cite{Garcia,Weinert}, but by 6 months of life daily rhythms in melatonin are apparent \cite{Kennaway} and consolidation of nighttime sleep has been established \cite{Bathory}. In toddlers and pre-school age children, melatonin rhythms are well-established \cite{LeBourgeoisJenni2013} reflecting a fully developed circadian system. Experimental studies of melatonin onset timing in toddlers indicate an earlier circadian phase compared to adults and adolescents \cite{LeBourgeoisJenni2013},  suggesting that the circadian propensity for wakefulness may also be earlier. Additionally, in pre-school age children melatonin rhythms show a higher sensitivity to evening light exposure compared to adults \cite{Hartstein2022}, suggesting that light processing in the circadian clock model may differ from that in adults. Thus, further experiments and modeling work are needed to refine circadian clock models for this age group. However, our results provide preliminary insights into potential physiological susceptibility for nighttime waking that can be contributing to this common sleep issue in young children.

Brief arousals during nighttime sleep, especially at transitions from rapid eye movement (REM) sleep to non-REM (NREM) sleep, are features of healthy human sleep. The HCL model does not explicitly account for different stages of sleep but instead models how the interactions of the homeostatic sleep drive and the circadian rhythm dictate the timing and duration of the propensity for sleep and wake behavior. While providing an incomplete picture of sleep behavior, these two processes and their interactions, as canonized in the classic Two-Process Model \cite{Daan1984,tpm}, have been shown to significantly influence sleep timing and duration, and can account for multiple other features of human sleep \cite{borbely}. Physiologically-based mathematical models for the network of sleep and wake promoting brain regions have been developed (reviewed in \cite{BoothDinizBehnreview,Postnovareview}). However, it has been shown that sleep-wake regulation network models that only include wake- and sleep-promoting populations in addition to  homeostatic sleep and circadian rhythm drives, known as sleep-wake flip-flop models, have equivalent solution dynamics to the Two-Process Model \cite{SkeldonTPMSWFF}. Thus, we expect that similar results could be obtained using a sleep-wake flip-flop model. An interesting question for future work is how our proposed susceptibility for nighttime waking based on circadian and homeostatic processes would interact with NREM-REM alternation during nighttime sleep and the effect on durations of normal post-REM brief awakenings.

In summary, our modeling results provide new understanding of how homeostatic and circadian processes can differ in young children and may be contributing to a common sleep disruption. Additionally, the model introduces a framework that can be used to explore external light interventions that can help consolidate and stabilize sleep behavior in young children.

\section{Model and Methods}

\label{sec:skeldon model}

The classic Two-Process Model represents human sleep and wake cycles due to the dynamic interplay between a circadian oscillator and an exponentially varying homeostatic sleep drive \cite{Daan1984, tpm}. Recently, Skeldon et al. \cite{skeldon2023method} introduced an extension of the Two-Process Model that incorporates a Van der Pol oscillator model \cite{forger1999simpler} for the circadian oscillator that can be entrained to an external light schedule. By fitting the circadian oscillator model to data from forced desynchrony studies, their Homeostatic-Circadian-Light (HCL) model accounts for modulations of the daily variation in sleep propensity in addition to dynamic effects of external light exposure. The following sections describe the components of the HCL model: the circadian rhythm and light processes, and the homeostatic sleep drive process.

\subsection{Circadian rhythm and light processes}
\label{sec:circadian}

As in the traditional Two-Process Model \cite{Daan1984, tpm}, the circadian rhythms of sleep and wake propensity in the HCL model are represented by upper and lower threshold curves, $H^+(t)$ and $H^-(t)$, respectively. These curves are defined as
\begin{align}\label{eq:H+}
    H^+(t) = H_0^+ + aC(t),\\
    H^-(t) = H_0^- + aC(t),
\end{align}
where $H_0^+$ and $H_0^-$ represent the baseline levels of the upper and lower thresholds, respectively, $a$ denotes the circadian amplitude and time $t$ is in days. The circadian rhythm $C(t)$ is based on the variables $c(t)$ and $x_c(t)$ of the Van der Pol oscillator model for the circadian clock in the suprachiasmatic nucleus (SCN) introduced by Forger et al. \cite{forger1999simpler}.  In the circadian clock model, $c$ typically represents the circadian drive output by the SCN and $x_c$ is a complementary variable. The governing equations for the circadian pacemaker (rescaled to time in days) are given by:
\begin{align} 
    \frac{dc}{dt} &= 2\pi (x_c + B), \label{eq:circadian c} \\
    \frac{dx_c}{dt} &= 2\pi \left[ \mu \left( x_c - \frac{4}{3} x_c^3 \right) - \left( \left( \frac{24}{f \tau_c} \right)^2 + kB \right)c \right]\label{eq:circadian x_c}.
\end{align}
External light is provided to the pacemaker through the functions $B(t)$, which generates the phasic and tonic effects of light on the circadian clock, and $\hat{B}(t)$, termed Process L which accounts for the effect of light on retinal photoreceptors:
\begin{align}
   B &= \hat{B} (1 - 0.4c)(1 - 0.4x_c), \\
   \hat{B} &= G_{\text{day}} (1 - n) \alpha.
\end{align}
The activation of $\hat{B}(t)$ is modulated by $n(t)$, which denotes the fraction of saturated photoreceptors in the retina. The dynamics of $n(t)$ are given by:
\begin{align} \label{eq:circadian n}
    \frac{dn}{dt} &= \alpha (1 - n) - \beta_{\text{day}} n,
\end{align}
The photoreceptor saturation rate $\alpha$  is given by $\alpha = \alpha_{0, \text{day}} \left( \frac{\tilde{I}(t)}{I_0} \right)^p$, where $\tilde{I}(t)$ denotes the external light level (in lux). Parameter values were fit to account for experimental data on human circadian phase responses to light \cite{forger1999simpler, forger2017biological}.

For the circadian rhythm $C(t)$ in the HCL model, Skeldon et al. fit a quadratic function of $c(t)$ and $x_c(t)$ to data from forced desynchrony experiments to account for daily modulation of the circadian drive for wakefulness:  
\begin{align} \label{eq:circadian C}
    C(t) = c_{20} + \alpha_{21}x_c + \alpha_{22}c + \beta_{21}x_c^2 + \beta_{22}cx_c + \beta_{23}c^2.
\end{align}

\subsection{Homeostatic sleep drive process}
\label{sec:Homeostatic}

As in the Two-Process Model, sleep pressure $H(t)$ is assumed to increase exponentially during wakefulness and decrease exponentially during sleep. During wakefulness, sleep pressure follows the differential equation:
\begin{align}
    \tau_{hw} \frac{dH_w}{dt} = -H_w + h_{\text{max}}, \quad \text{with initial condition} \quad H_w(t_w),
\end{align}
where $\tau_{hw}$ (in days) is the time constant governing the rate of increase, $h_{\text{max}}$ (in \%SWA) is the upper asymptote of sleep pressure, and $t_w$ is the time of wake onset.

Similarly, during sleep, sleep pressure decays exponentially according to:
\begin{align}
    \tau_{hs} \frac{dH_s}{dt} = -H_s + h_{\text{min}}, \quad \text{with initial condition} \quad H_s(t_s),
\end{align}
where $\tau_{hs}$ (in days) is the time constant governing the rate of decrease, $h_{\text{min}}$ (in \%SWA) is the lower asymptote of sleep pressure, and $t_s$ is the time of sleep onset.

The analytical solutions to these differential equations are given by:
\begin{align}
    H_w(t) &= h_{\text{max}} - \left(h_{\text{max}} - H_w(t_w)\right) \exp\left(-\frac{t - t_0}{\tau_{hw}}\right),\label{eq:H_w} \\
    H_s(t) &= h_{\text{min}} - \left(h_{\text{min}} - H_s(t_s)\right) \exp\left(\frac{t - t_0}{\tau_{hs}}\right),\label{eq:H_s}
\end{align}
Transitions between wakefulness and sleep are dictated by the circadian rhythm threshold curves $H^+(t)$ and $H^-(t)$ (Eq. \ref{eq:H+}). A switch from wake to sleep occurs when $ H_w(t_s) = H^+(t_s)$, 
and a switch from sleep to wake occurs when $H_s(t_w) = H^-(t_w)$.

\subsection{Parameter fitting}
\label{sec:parameters data}

To simulate sleep patterns in young children, we restrict the parameters governing the homeostatic sleep drive to experimentally identified ranges from sleep studies of healthy 2.5-3 year olds \cite{lebourgeois2012dynamics}. The experimentally estimated values for the  homeostatic rise and decay time constants, $\tau_{hw}$ and $\tau_{hs}$, respectively, and for the upper and lower asymptotes, $h_{\max}$ and $h_{\min}$, are listed in Table \ref{tab:data}. Setting the parameters for the circadian rhythm and light processes to values estimated for adults as in \cite{skeldon2023method} (Table \ref{tab:parameters}), we used a differential evolution algorithm \cite{buehren2025differential} to determine the values for the homeostatic sleep drive parameters, the circadian amplitude $a$ and the baseline levels of the upper and lower circadian thresholds, $H^+_0$ and $H^-_0$, respectively, to generate a typical habitual sleep pattern measured in healthy children of ages $34.2 \pm 2$ months \cite{Akacem2015,athanasouli2024data} (Table \ref{tab:data}). With these optimized parameter values (Table \ref{tab:optimized}, Optimized values), the simulated sleep pattern consists of a daytime nap and a nighttime sleep episode at the mean times and durations shown in Table \ref{tab:data}. For this parameter optimization, the external light schedule was set to a default schedule that matched the experimental wake and bed times with a dim light period based on nap midpoint time \cite{athanasouli2024data}:  bright light (1000 lux) begins at 06:58, dim light (10 lux) starts at 14:00 and ends at 15:30, and lights are extinguished (0 lux) at 20:18.
The default evening lights-off time was based on the reported bed time \cite{Akacem2015} which is earlier than the reported sleep onset time included in Table \ref{tab:data}.

\begin{table}[h!]
\centering
\begin{tabular}{|c|c||c|c|}
\toprule
\textbf{Parameter} & \textbf{Value} 
  & \textbf{Parameter} & \textbf{Value}  \\
\midrule
$\tau_c$ (day)            & 1.0083         &  $I_0$ (lux)& 9500  \\
$f$                       & 0.99669        &  $c_{20}$                & 0.7896  \\
$G_{\mathrm{day}}$ (day)& 0.0138           &  $\alpha_{21}$           & -0.3912  \\
$p$                       & 0.6            &  $\alpha_{22}$           & 0.7583     \\
$k$                       & 0.55           &  $\beta_{21}$            & -0.4442       \\
$b$                       & 0.4            &  $\beta_{22}$            & 0.0250         \\
$\alpha_{0,\mathrm{day}}$ (day$^{-1}$)& 230.4 & $\beta_{23}$            & -0.9647   \\
$\beta_{\mathrm{day}}$ (day$^{-1}$)& 18.72 &    &             \\
\bottomrule
\end{tabular}
\caption{Parameter values for the circadian and light processes of the HCL model (Equations \ref{eq:circadian c}, \ref{eq:circadian x_c}, \ref{eq:circadian n}, \ref{eq:circadian C}) \cite{skeldon2023method}.}
\label{tab:parameters}
\end{table}

To explore ranges of parameter values that replicate the experimentally observed sleep patterns in young children, we surveyed parameter values using Latin hypercube sampling under several key assumptions. First, we constrained the homeostatic parameters $\tau_{hw}$, $\tau_{hs}$, $h_{\max}$, and $h_{\min}$ to lie strictly within the experimentally observed ranges reported in Table \ref{tab:data}. For the varied circadian parameters, we assumed the amplitude $a \in [0, 300]$ and set the baseline levels of the upper and lower circadian thresholds, $H^+_0$ and $H^-_0$, respectively, within $[0, 400]$. These intervals were chosen to span the full dynamic range from the minimum of $h_{\min}$ to the maximum of $h_{\max}$, while ensuring that the lower threshold remained strictly positive. Under these constraints, we used Latin Hypercube Sampling to generate a well-distributed set of 100,000 parameter samples and identified those sets that produced simulated sleep metrics within the empirical bounds specified in Table \ref{tab:data}.  Since the union of viable parameter combinations forms an irregular shape in 7-dimensional parameter space, we identified the smallest axis-aligned hypercube that fully encloses the admissible parameter combinations. The resulting feasible parameter ranges are summarized in Table \ref{tab:optimized}. 

\renewcommand{\abstractname}{Acknowledgments}
\begin{abstract}
T.Y. was supported in part by the University of Michigan Department of Mathematics Research Experience for Undergraduates (REU) program. 
\end{abstract}

\renewcommand{\abstractname}{Statement on Code Availability}
\begin{abstract}
\sloppy{All code used to generate figures is archived on GitHub ( https://github.com/YuzuruTesla/YaoBooth-TwoProcessModel) and licensed for reuse, with appropriate attribution/citation,
under a BSD 3-Clause Revised License. This repository contains the Matlab code used to run simulations and to generate the figures in the paper.}
\end{abstract}

\bibliographystyle{unsrt}
\bibliography{References}

\end{document}